\documentclass[twocolumn,pre,preprintnumbers,superscriptaddress]{revtex4}
\usepackage{textcomp}
\usepackage{amsmath}
\usepackage{amssymb}
\usepackage{graphicx}
\usepackage{color}
\usepackage{soul}
\newcommand{\be}{\begin{equation}}
\newcommand{\ee}{\end{equation}}
\newcommand{\ba}{\begin{eqnarray}}
\newcommand{\ea}{\end{eqnarray}}

\newcommand{\cmp}
{\affiliation{Saha Institute of Nuclear Physics, Kolkata 700064, India.}}
\newcommand{\barasat}
{\affiliation{Barasat Government College, Kolkata 700124, India.}}

\begin{document}

\title{Quantum Annealing and Computation}

 \author{Bikas K Chakrabarti}
 \email{bikask.chakrabarti@saha.ac.in}
 \cmp

 \author{Sudip Mukherjee\footnote{Corresponding author}}
 \email{sudip.mukherjee@saha.ac.in}
 \barasat \cmp

 \begin{abstract}
We introduce and review briefly the phenomenon of quantum annealing and
analog computation. The role of quantum fluctuation (tunneling) in random
systems with rugged (free) energy landscapes having macroscopic barriers
are discussed to demonstrate the quantum advantage in the search for the
ground state(s) through annealing. Quantum annealing as a physical
(analog) process to search for the optimal solutions of computationally
hard problems  are also discussed.\\


 \end{abstract}
 \maketitle
 
\section{Introduction} 
The idea of computation by decomposing the entire operation into elementary operations like addition, subtraction, etc, and eventually 
recasting the problem into just number crunching arithmetic, has been quite successful and popular. This has been possible because 
of the feasibility of binary digitization  of the real numbers and availability of ever faster electronic logic gates (AND, OR, XOR, 
NOT, etc.).

The  development  and  success  of digital  computers eclipsed the development of the analog computers, based on the dynamics of 
physical systems. Imagine a bowl on the table and you need to ‘locate’  its  bottom point.  Of  course,  one  can  calculate
the local  depths  (from  a  reference height)  everywhere  along the inner surface of the bowl and search for the point where the 
local depth is maximum.
However, as every one would easily guess, a  much  simpler  and  practical  method could be to allow rolling of a marble ball along 
the inner surface of the bowl and wait for its resting location or position. Here, the physics of the forces of gravity and friction 
allows us to ‘calculate’ the location of the bottom-most (or minimum energy) point in an analog way! In principle, a similar trick 
would work for cases where the internal surface of the bowl has modulations, keeping the surface contour or the landscape valleys   
all tilted towards the same bottom point location. Problem comes when these valleys get separated by  `barriers',  high enough.

Computationally hard problems, e.g., search of the ground state(s) of $N$-spin Sherrington-Kirkpatrick (SK) spin glass system~\cite{sudip-sk} 
involves locating the minima in a rugged landscape of size $2^N$, or exponential in $N$. The (free) energy landscape becomes rugged with 
occasional barriers of $N$ order, meaning that by flipping a countable few spins one can not get out of the local minimum. For that, one needs 
to flip a macroscopic number (of order $N$) to get out of the local minima. The same is true for the $N$-city Traveling Salesman Problem 
(see e.g.,~\cite{sudip-SA}) where one has to find the minimum cost (or travel distance) tours for visiting all the cities, from among the 
$N!$ order possible tours. Here also, perturbing a tour of higher travel length
(cost function), by rearranging the local visit sequence
(of a few cities), will not lead to success and one has to take a global view of modifying the sequence of visits to a finite fraction of the 
cities (effectively by overcoming $N$ order cost function barriers). Generally, for such minimum cost search from among $\exp(N)$ or higher configurations (or
tours), often separated by $N$-order (energy or cost function) barriers, the search time can not be bounded by any polynomial in $N$ (NP-hard problems).

\section{Simulated (classical) annealing}
In the seminal paper~\cite{sudip-SA} by Kirkpatrick
et al., a novel stochastic technique was
proposed based on  the  metallurgical
annealing technique:  To search for the
optimized cost (energy  of  the  ground
state  `crystal')  at  eventually vanishing
noise  (or  temperature $T$),  one  starts
from  a  high noise ($T$) `melt' phase, and
tune slowly the noise level.  In  this
‘simulated noise’ tuning process,  the
(classical)  noise  at any  intermediate
level  of  annealing  allows   for  the
acceptance with (Gibbs-like) probability
determined by the change  in  cost or
energy (scaled  by the  noise factor)  of
even  higher  cost  (energy) fluctuations.
As the noise level ($T$) is  slowly reduced
during the  annealing,  the  gradually
decreasing  probability  of accepting
higher  cost  values  allows  the  system
to  come out of the local minima valleys
and to settle eventually in the ‘ground  state’
of  the  system  with  lowest  cost  (energy)
value. It has been a remarkably successful
trick for ‘practical’ computational solutions
of a large class of  multi-variable  optimization
problems in reasonable convergence time. For
NP-hard problems, however, the  search or convergence
time  for reaching  the  lowest  cost  state
or configuration grows  exponentially with $N$.

The  bottleneck  could  be  identified  soon. Extensive study  of  the  dynamics  of frustrated  random  systems  like the 
SK model  showed that its (free) energy landscape (in the spin glass phase) is extremely rugged and the barriers, separating 
the local valleys, often become $N$ order for a $N$-spin glass. Same is true for $N$-city Traveling Salesman Problem. In the 
macroscopic size limit ($N$ approaching infinity) therefore such systems become (effectively) non-ergodic or localized and 
the classical (thermal) fluctuations, like that in the simulated annealing, fail to help 
the  system  to come out  of  such  high barriers  (at  random locations  or configurations, not  dictated  by  any symmetry) as the 
escape  probability  is of  order  $\exp(-N/T)$  only. Naturally the annealing time (inversely proportional to the escape probability), 
to get the ground  state  of  the  $N$-spin SK model, can not be bounded by any polynomial in $N$.

\begin{figure}[htb]
\begin{center}
\includegraphics[width=7.0cm]{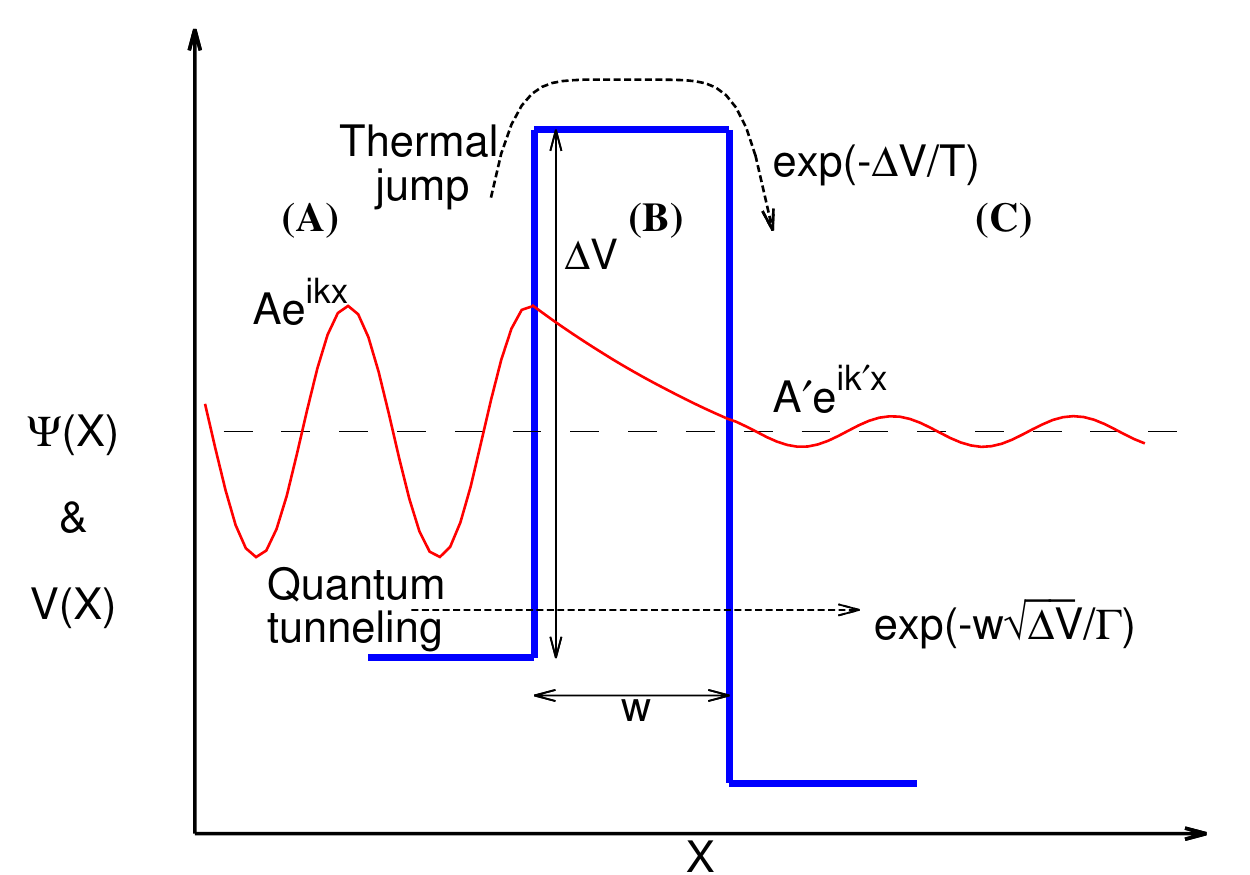}
\end{center}
\caption{Schematic diagram suggesting the advantage of
quantum tunneling over thermal hopping across
a potential barrier. For kinetic energy ($E$)
greater than the barrier potential $(V)$,
compared to those in regions (A) and (C), the
plane wave functions ($\psi(x) = A\exp(ikx)$
or $\psi(x) = A'\exp(ik'x)$)  are indicated
with (real) wave vectors $k = \sqrt{E - V} \simeq k'$
in the two regions respectively. For region (B),
the wave vector becomes imaginary (as $V > E$)
and the consequent damping of the amplitude $A$
in region (A) to $A'$ in region (C), occurs while 
passing through region (B). For $V$ of order $N$
and $E$ of order $N\Gamma$, the barrier
transmission probability $(A'/A)^2$ scales as
$\exp(-w\sqrt{N (1 - \Gamma)}) \sim
exp(-w\sqrt{N}/\Gamma)$.}
\label{area_ergodic}
\end{figure}

\begin{figure}[htb]
\begin{center}
\includegraphics[width=7.0cm]{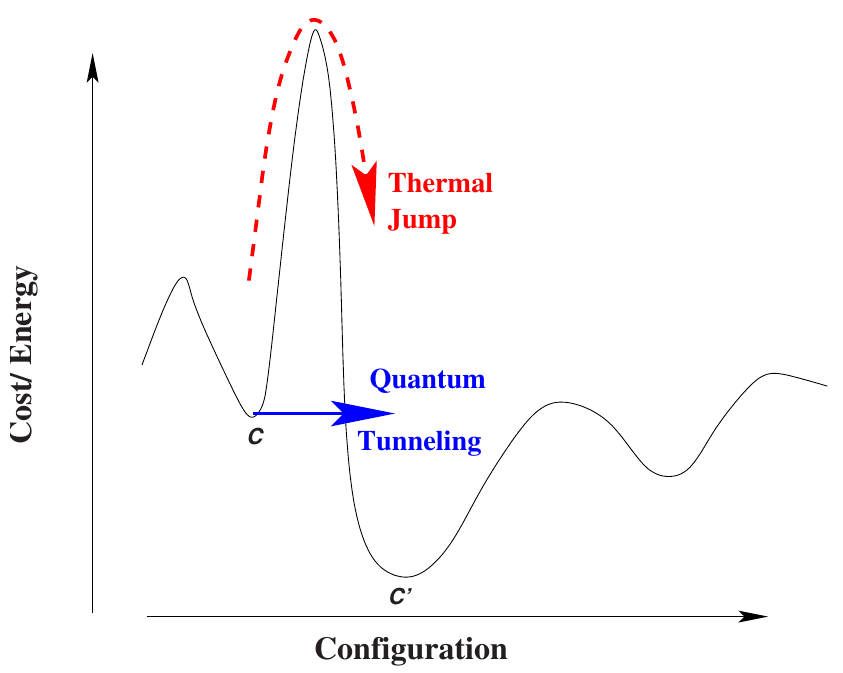}
\end{center}
\caption{Optimizing the cost function of a computationally hard problem
(for example, search for the ground state energy of a spin glass or for
the minimum travel distance in  a traveling salesman problem), one has to
get out of a shallower local minimum (indicated here by the configuration 
$C$ of the spins in a spin glass or of a travel route of the
salesman), to reach a deeper minimum (indicated here by $C'$). This
requires jumps or tunneling like fluctuations in the dynamics. Classically
one has to jump over the energy or the cost barriers separating them,
while (as indicated in Fig.~\ref{area_ergodic}) quantum mechanically one can tunnel through
the same. With increasing barrier  height, thermal jump becomes extremely
difficult. However, if the barrier is narrow enough, quantum tunneling can
help the search problem much easily.}
\label{rmp_fig}
\end{figure}

\section{Quantum tunneling advantage for annealing}
The indication~\cite{sudip-ray} was that quantum
fluctuations in the SK model can perhaps
lead to some escape routes to ergodicity
or quantum fluctuation  induced
delocalization  (at  least   in  low
temperature  region  of  the  spin  glass
phase)  by  allowing tunneling through
such macroscopically tall but thin barriers
which are difficult to scale using classical
fluctuations. This is  based  on  the
argument that  escape  probability (see
e.g., the discussion in ref.~\cite{sudip-bikas})
due   to quantum  tunneling,  from  a
single valley  with  single  barrier of
height $V$ and width $w$, scales as
$\exp (- \sqrt{V} w/\Gamma)$; where  $\Gamma$
represents the  quantum fluctuation
strength  (or tunneling probability),
while the classical  (escape)
probability is of  order  $\exp (-V/T)$
with $T$ denoting the temperature
of the system (see Fig.~\ref{area_ergodic}). This extra handle
through the barrier width $w$ (absent in
the  classical  escape  probability)
helps in a major way in its vanishing
limit. Indeed, for a single narrow
($w \rightarrow$  0)  barrier  of
height $N$,  when the tunneling parameter
$\Gamma$ is  slowly  tuned to zero,  the
annealing  time  to  reach  the  ground
state  or optimized  cost,  will  become
$N$  independent!  It  has  led  to some
important clues. Of course, complications
(due to localization)  may  still  arise
for  many  such  barriers  at random
locations.

In the Quantum Annealing scheme, the cost
function of a multivariable optimization problem
is mapped on to the energy function corresponding
to a classical (frustrated) Hamiltonian ($H_0$).
A time dependent quantum (non-commuting) kinetic term
($\Gamma (t)H'$) is then added to the system. As
$H'$ does not commute with the $H_0$, it provides
quantum dynamics to the overall system represented
by the time dependent Hamiltonian $H(t) = H_0 +
\Gamma(t)H'$ and the evolution of the system is
given by the solution of the time dependent
Schrödinger equation of the system

\begin{align*}
i{\hbar}{\frac{\partial \psi}{\partial t}} = \Big[ \Gamma (t)H'+H^0\Big]\psi .
\end{align*}

\begin{figure}[htb]
\begin{center}
\includegraphics[width=7.0cm]{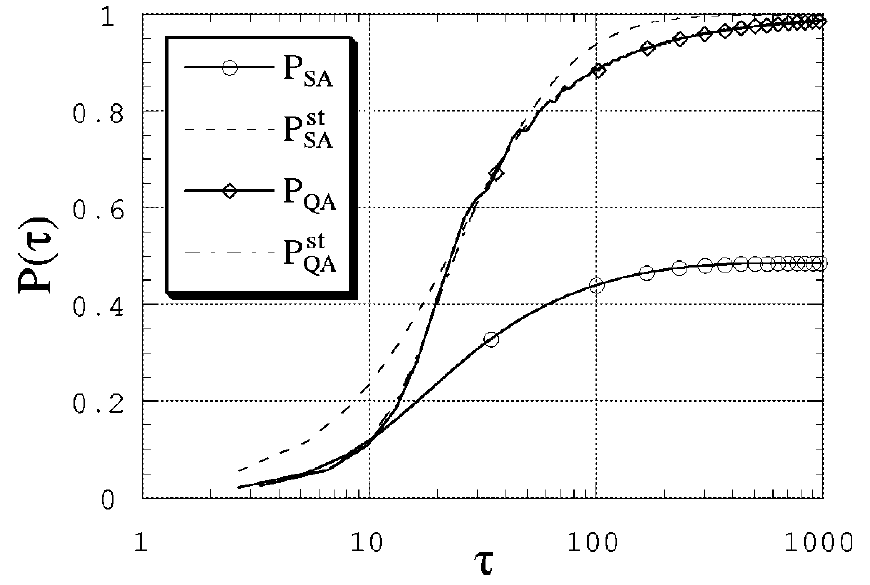}
\end{center}
\caption{Time dependence of the
overlaps $P$ (with the ground state)
of the frustrated model with
$\Gamma (t) = T(t) = 3/{\sqrt t}$ for
classical and quantum annealing
(subscripts `SA' and `QA' indicate simulated annealing and quantum annealing respectively, whereas 
superscript `st' corresponds to 
quasistatic solution). [Fig. taken from ref.~\cite{sudip-Nishimori}]}
\label{nishi_QA}
\end{figure}

\noindent If $\Gamma(0)$ is taken to be very large, then
$\psi$ starts as the ground state of $H'$,
which is assumed to be known. As $\Gamma(t)$ starts decreasing slowly enough,
following the quantum adiabatic theorem, the
system will then be carried into 
the ground state of the instantaneous
total Hamiltonian. At the end of the annealing
schedule the kinetic term becomes zero ($\Gamma = 0)$.
Hence, one would expect the system to arrive
at (one of) the ground state(s) of $H_0$, thereby
giving the optimized value of the original cost
function. We represent schematically the
advantage of quantum annealing, compared to the classical one, using Fig.~\ref{rmp_fig}. 
This was first demonstrated~\cite{sudip-Nishimori} 
for random spin systems, by numerical solutions of the above 
Schrodinger equation for such systems. Indeed, the numerical 
results reported there for the annealing of a frustrated
magnetic system (with less than ten spins),  
are shown in Fig.~\ref{nishi_QA}, indicating the clear
advantage of tuning the quantum fluctuation, compared to the
classical (thermal) one. The successive theoretical~\cite{sudip-Farhi,sudip-Santoro} and
experimental~\cite{sudip-Brooke} demonstrations led to the emergence of the field~\cite{sudip-das}.

\begin{figure}[htb]
\begin{center}
\includegraphics[width=6.3cm]{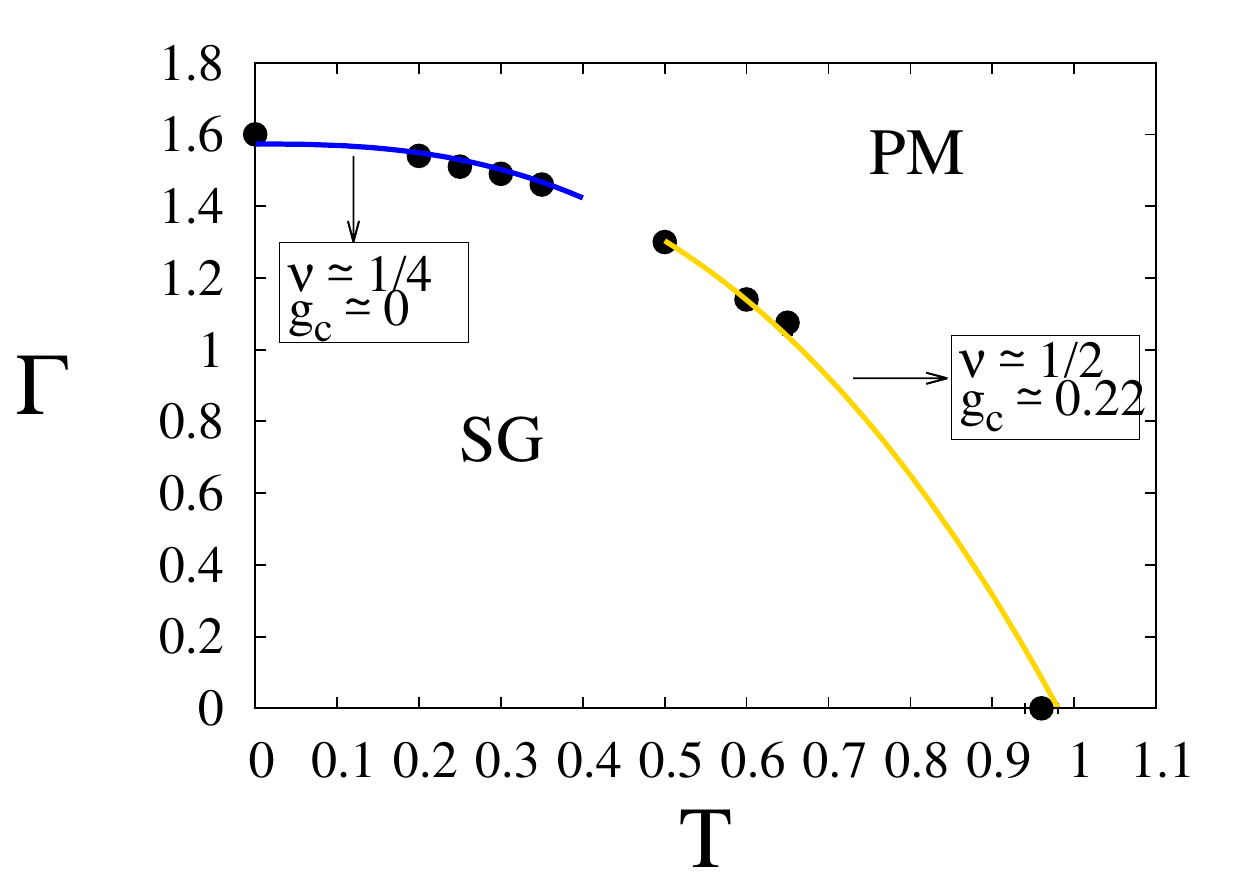}
\end{center}
\caption{Phase diagram of quantum SK spin
glass model as estimated from the Monte Carlo
simulation and exact diagonalization.  The spin
glass and paramagnetic phases are denoted by SG
and PM respectively. The data points at $T = 0$
and $\Gamma = 0$ are associated with the purely
quantum and classical phase transitions,
respectively. The two observed different critical
behaviors of the system are indicated by blue
(with the Binder cumulant $g_c$ near $0$ and the correlation
length exponent value near $1/4$) and by yellow (with critical
Binder cumulant $g_c$ near $0.22$ and correlation length
exponent value near $1/2$) lines. The crossover in the critical behavior
occurs around $T = 0.49$  and $\Gamma = 1.31$. 
[Fig. taken from ref.~\cite{sudip-jsps}]}
\label{phase_diagram}
\end{figure}

\section{Quantum annealing in spin glasses}
Using Suzuki-Trotter mapping of the quantum SK model to the effective
classical spin model, Monte Carlo studies were conducted and
annealing through ergodic and nonergodic
regions of the model was studied~\cite{sudip-jsps}. 
Determination of thee phase diagram of the quantum
SK model, employing the Monte Carlo simulation (at
finite temperatures) and exact diagonalization
technique (at zero temperature) revealed some
interesting features.  To extract the critical
behavior at finite temperatures, different sizes
(number of Ising spins ranging from $20$ to $180$)
were considered in~\cite{sudip-jsps}, keeping constant the
ratio of Trotter size and the effective dynamical
size of the number of spins.  At zero temperature,
however, the  maximum system size (for diagonalization study)
had been of order 20 only.
They found that from the quantum transition point
($T = 0, \Gamma_c \simeq 1.63$) to almost the
point ($T = 0.45, \Gamma = 1.33$), the critical
Binder cumulant value  remained vanishingly small
(it can indeed effectively vanish  even for 
non-Gaussian fluctuation-induced phase transitions).
In this range of the phase boundary, they found
the correlation length exponent to be about $1/4$
from the data collapse of Binder cumulant plots.
In the rest of the phase boundary, the critical
Binder cumulant value is about $0.22$  and they
observed a satisfactory data collapse with the
correlation length exponent equal to $1/2$. These
two different values of the critical Binder
comulant and the correlation length exponent
for the two different parts of the phase
boundary indicate the classical to quantum
crossover (at $T \simeq 0.49$ and $\Gamma \simeq
1.31$; see Fig.~\ref{phase_diagram}) occurs at a non-vanishing value of temperature in the quantum SK model.

\begin{figure}[htb]
\begin{center}
\includegraphics[width=6.7cm]{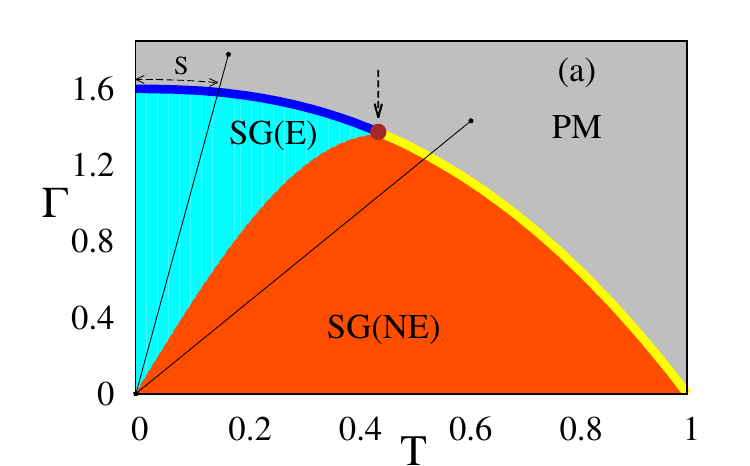}
\includegraphics[width=6.8cm]{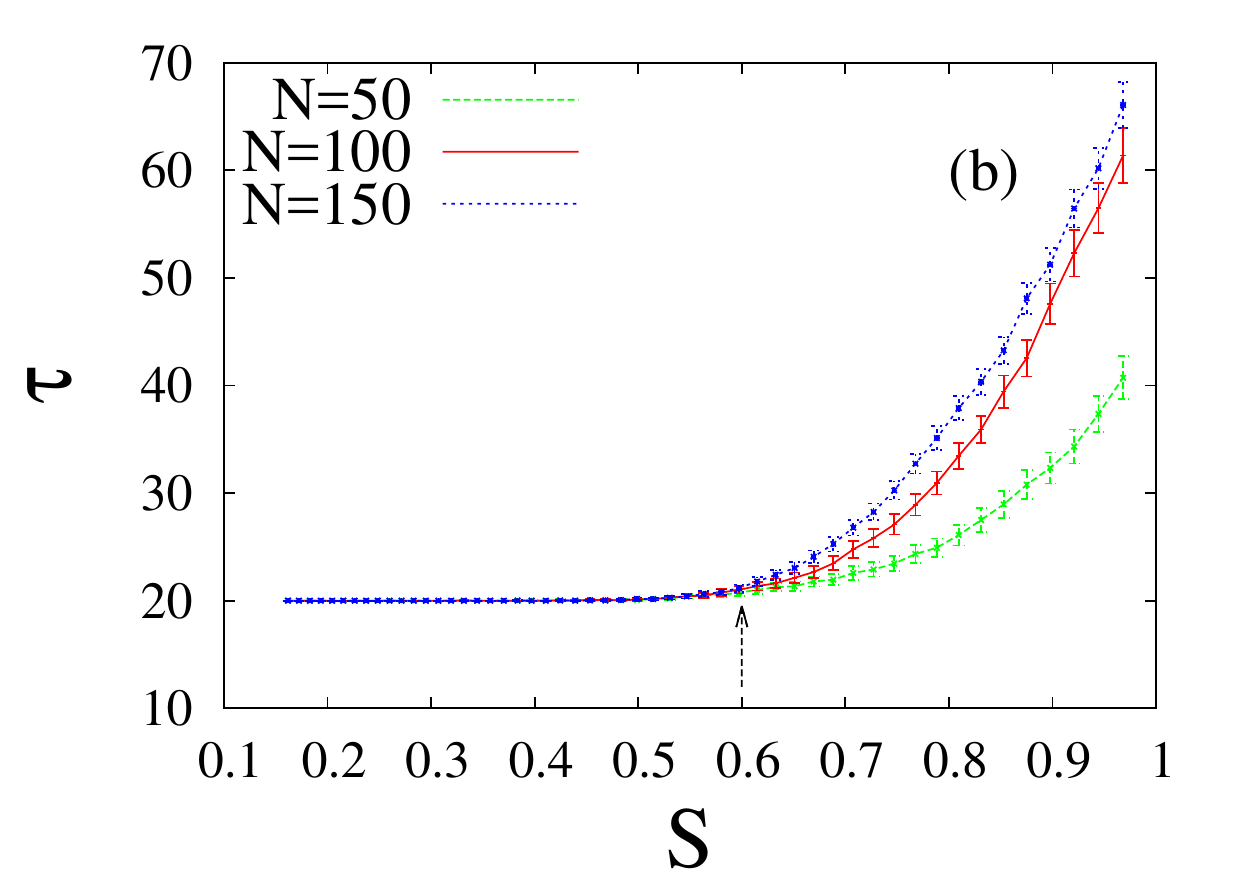}
\end{center}
\caption{(a) Schematic phase diagram of the
quantum SK model. The spin glass and paramagnetic
phases are denoted by SG and PM, respectively.
Numerical results showed~\cite{sudip-jsps} that the spin
glass phase is further  divided into two regions:
ergodic spin glass phase SG(E) and nonergodic spin glass
phase SG(NE). The quantum-classical crossover point
in the critical behavior of the model is indicated by
the red dot on the SG-PM phase boundary. The
annealing behavior for the schedules 
$T(t) = T_0 (1 - t/\tau)$ and $\Gamma(t) = \Gamma_0 (1 - t/\tau)$, with $T_0$ and $\Gamma_0$ values corresponding
to the paramagnetic phase are shown in (b). The
Variation of annealing time $\tau$ with S, the length
of the arc measured along the phase boundary (starting
from the zero-temperature quantum transition point)
to the intersection of the annealing line with the
phase boundary. Low values of $S$ correspond to
the SG(E)-PM phase boundary, while higher values of
$S$ correspond to SG(NE)-PM boundary. The annealing
time $\tau$ is clearly system size independent  when
the annealing path passes through the SG(E) region
and increases rapidly with the system size when it
pass through the SG(NE) region. 
[Fig. taken from ref.~\cite{sudip-jsps}]}
\label{ergodic}
\end{figure}

Unlike in the pure system (also in non-frustrated random system),  where the
free-energy landscape  is smoothly inclined
towards the global minima (Landau scenario), in
the SK spin glass the landscape is extremely
rugged. In particular, the local minima are often
separated by system size dependent  energy
barriers which induce  nonergodicity and the
consequent replica-symmetry-broken distribution
of the order parameter. Therefore,  at any
finite temperature the thermal fluctuations
are unable to help the localized system to come
out or escape from the macroscopically high
free-energy barriers to reach the ground state
(by flipping finite fraction of spins). With the
aid of the transverse field the system can
tunnel through such free-energy barriers~\cite{sudip-ray,sudip-bikas,sudip-das}. 
As a consequence, at low temperatures, the
phase transition is governed by the quantum
fluctuation and the system essentially exhibits
quantum critical behavior (see the discussion
on ``possible restoration of ergodicity through quantum
tunneling'' for appropriate parameter space in
the quantum SK model~\cite{sudip-Leschke}).

Study on the nature of the order parameter
distribution in the spin glass phase at finite
temperatures through Monte Carlo simulations also
showed~\cite{sudip-jsps} that in the high-temperature
(low-transverse-field) classical fluctuation
dominated spin glass region, along with the peak
at the most probable value of the order parameter,
the distribution contains a long tail (extending
up to the zero value of the order parameter). This
tail did not vanish even in the large system size
limit, which indicated  that the order parameter
distribution remains Parisi type (replica Symmetry
broken), corresponding to the nonergodic region
SG(NE) (Fig.~\ref{ergodic}a)
of the spin glass phase. On the other hand, in the
low temperature (large transverse field) region,
where the order parameter distribution effectively
converges to a Gaussian form (with a peak around
the most probable value) in the infinite-system-size
limit. This indicates the existence of a single
(replica-symmetric) order parameter in this ergodic
region SG(E) of the spin glass phase. At zero
temperature, the extrapolated order parameter
distribution function showed a clear tendency to
become one with a sharp peak (around the most
probable value) in the large-system-size limit,
suggesting the conclusion that  the ergodic and
non-ergodic regions of the spin glass phase
are separated by a line possibly originating
from point ($T = 0, \Gamma = 0$) and extending up
to the quantum-classical crossover point ($T$
near  $0.49$ and $\Gamma$ near $1.31$) on the phase
boundary (see Fig.~\ref{ergodic}a). In order to find the
role of such quantum-fluctuation-induced ergodicity
in the (annealing) dynamics, they studied the
variation of the annealing time $\tau$ (re-
quired to reach close to the ground state(s)) with
the system size following the schedules $T (t) =
T_0 (1 - t/\tau)$ and $\Gamma(t) = \Gamma_0
(1 - t/\tau$). Attempts were made to reach a desired
preassigned very low energy state (near the
ground state) at the end of the annealing dynamics
(in time $\tau$), keeping both $T$ and $\Gamma$
nonzero (but very small) at the end of the
annealing schedule as the Suzuki-Trotter
Hamiltonian (which governed the annealing
dynamics) has singularities at both $T = 0$
and $\Gamma = 0$ (starting with $ T_0$ and
$\Gamma_0$ corresponding to the paramagnetic
region of the phase diagram). They found
(see Fig.~\ref{ergodic}b) that the average annealing time
does not depend on the system size when annealing
is carried out along paths that pass entirely
through the ergodic region, whereas the annealing
time becomes much larger and strongly size-dependent
for paths that pass through the nonergodic region
of the spin glass phase. These  confirm the earlier
observations~\cite{sudip-ray,sudip-bikas} regarding the annealing time
behavior, occurring due to tunneling through
macroscopically tall but thin free-energy barriers
in the SK model. In fact, the temporal variations
of the average spin autocorrelation  at finite
temperatures also showed distinct behavior in these
two regions of the spin glass phase of the quantum
SK model. While at high temperatures, in the classical
fluctuation dominated (nonergodic) region of the spin
glass phase (SG(NE) region in Fig.~\ref{ergodic}a), one obtains
very large values of the fitting relaxation times,
the fitting relaxation times become order of magnitude
smaller in the low temperature quantum fluctuation
dominated region (SG(E) in Fig.~\ref{ergodic}a) presumably because
of quantum tunneling~\cite{sudip-jsps}.
These investigations clearly indicated the existence
of a high-temperature 
(low-transverse-field) nonergodic region as well as a
low-temperature (high-transverse-field) ergodic
region in the spin glass phase. The line
separating these two regions starts from
$T = 0$, $\Gamma = 0$ and intersects the spin
glass phase boundary at the quantum-classical
crossover point (see Fig.~\ref{ergodic}(a)). The annealing
behaviors are indicated in Fig.~\ref{ergodic}(b), showing that the 
annealing time $\tau$ is practically independent of $N$ when the 
annealing schedule (line) passes through the ergodic spin glass region (at low $T$ and high $\Gamma$).

\begin{figure}[h]
\begin{center}
\includegraphics[width=6.5cm]{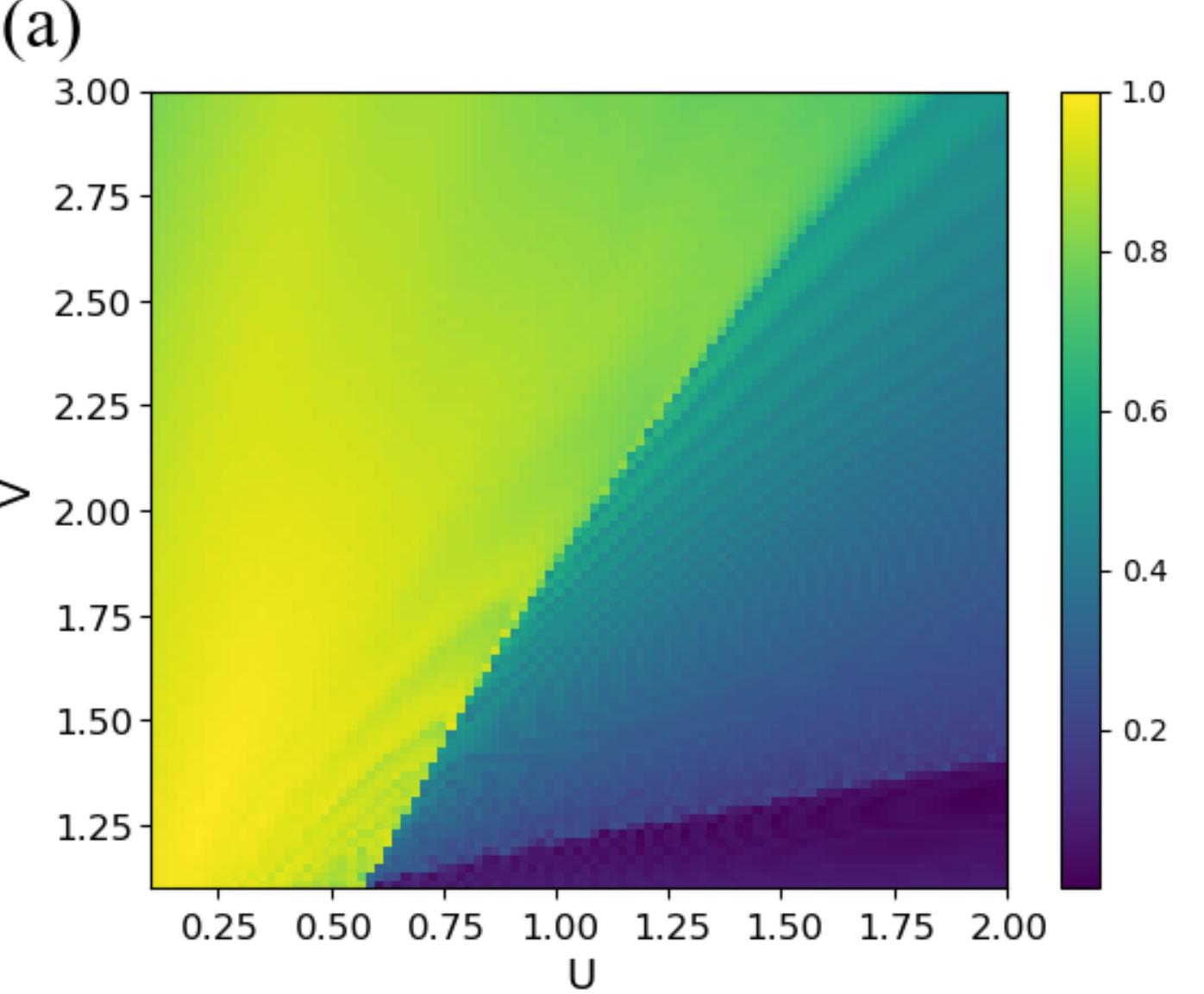}
\includegraphics[width=6.7cm]{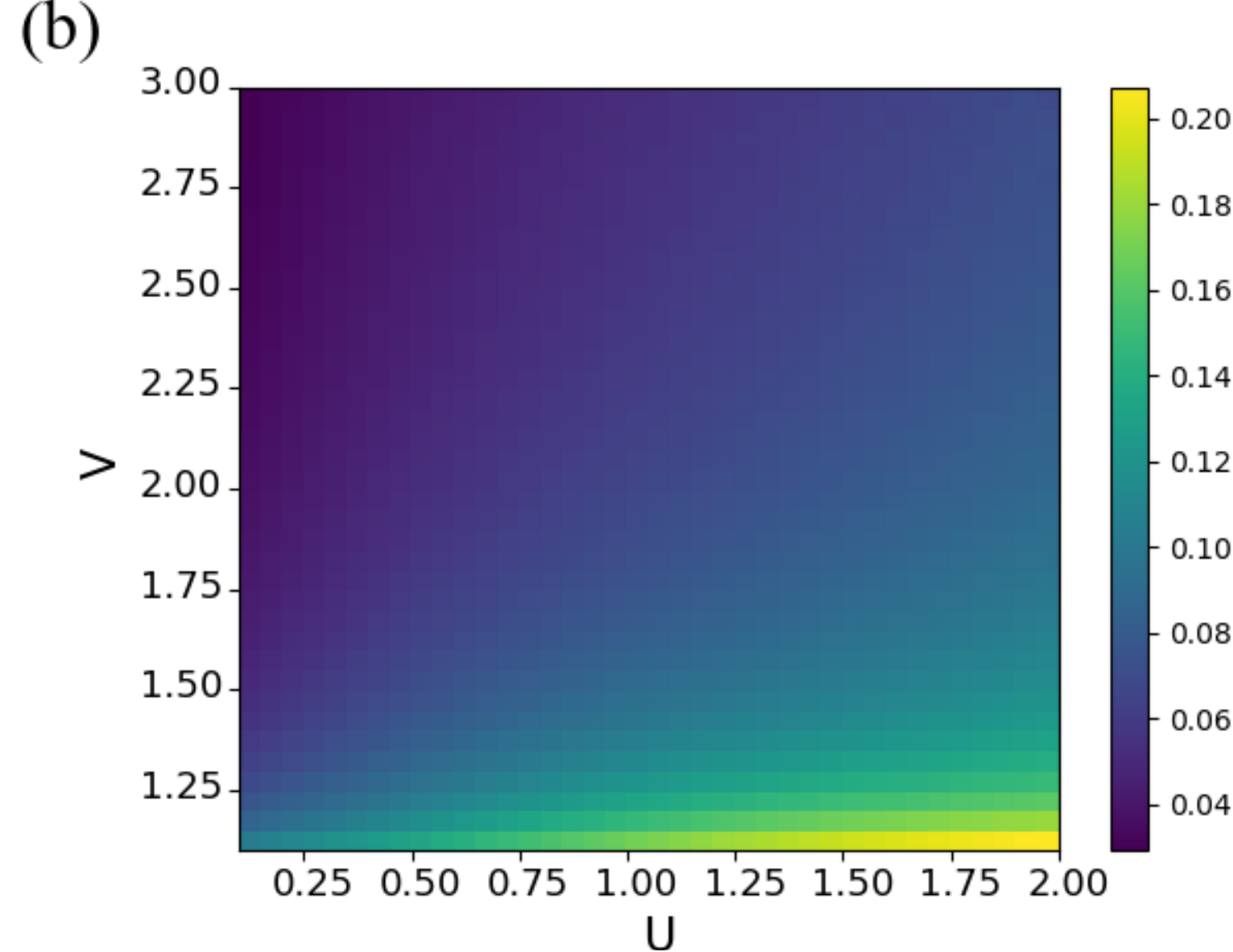}
\end{center}
\caption{(a) Color density plot showing the solution
fidelity after an adiabatic sweep using a
Hamiltonian description with a classical
mean-field approximation of the photon cavity
modes. The parameter $U$ describes the
particle-particle on-site repulsive
interaction and the parameter $V$ represents
the total potential depth of the specifically
shifted lattice site, to create an impurity
with a unique optimal solution.  Three regions
are clearly visible, a green-yellowish one,
where the correct solution of two-particle on
the mentioned lattice site is found, a bluish
region with finite overlap but non-optimal
solution, and a dark purple region with
minimal overlap to the optimal solution.
(b) Color density plot for the same parameter
range showing the maximum entropy generated
during the adiabatic sweep. In comparison, a
connection between low maximum entropy and
high success probability can be seen [Courtesy,
E. Starchl~\cite{sudip-Stachel}; private communication]}
\label{entropy_plot}
\end{figure}

\section{Quantum annealer using optical lattice in a cavity}
Indeed, Starchl and Ritsch~\cite{sudip-Stachel} studied a minimal Bose-Hubbard type system containing
two atoms inside a cavity-generated optical lattice induced by an external trapping potential and directly 
pumped by two laser beams resulting in  photon scattering into two separate cavity modes. Introduction
of an  additional local energy shift for a specific lattice site allows creating a unique energy separated 
optimal ground state for the problem Hamiltonian. The main focus was to compare the full quantum mechanical 
model with the semi-classical model, where the field mode induced interactions are approximated by classical
coherent fields. The investigation revealed strong differences between the quantum and semi-classical model, 
which can be cast into a sort of phase diagram separating a distinct parameter region where (partial) quantum 
annealing does not find the optimum in the semi-classical approach, while  full quantum annealing does. They 
found strong evidence for the involvement of entanglement in two ways. First, they~\cite{sudip-Stachel} confirmed  
the importance of vanishing entanglement at the end of the protocol and its influence on the minimal time needed 
for a successful simulation. Second, they identified a novel role of the maximum entanglement generated during the 
quantum annealing protocol in finding an optimal solution to the problem. They also studied the impact of different 
photon number cut-offs in the numerical quantum simulation. Most surprisingly they found improvement in the 
optimization success rate even for short simulation times when they used too low cut-off numbers to get a numerically 
accurate solution of the Schrödinger equation. Using higher cut-off numbers and thus a better representation of the 
exact quantum dynamics seems to be more relevant only for longer simulation times with very high success
rates (see Fig.~\ref{entropy_plot}). These observations indicate a clear quantitative improvement of the annealing process for a full quantum system compared to a semi-classical one. 

\section{Outlook  and further reading}
Quantum Annealing is now a very well developed subject, both theoretically as well as experimentally 
(see e.g.,~\cite{sudip-das,sudip-Brooke}). The breakthrough implementation of quantum annealing, employing 
superconducting qubits in successive generations of the D-wave annealers~\cite{sudip-Johnson} since 2011 and their 
availability in the market, has helped the growth of analog quantum computing in a major way. In fact, the availability of 
commercial quantum annealing condensed matter devices (like the D-wave systems)
have revolutionized quantum information processing and computing algorithms for optimization 
across many disciplines, and led to developments which were unthinkable about a decade back. We would refer 
the readers to a few recent studies to get further informations and ideas. For discussions on quantum versus classical 
annealing the readers may consult ref.~\cite{sudip-Heim}, while for discussions on the efficiency of quantum versus classical
annealing in the context of learning problems, the readers may consult ref.~\cite{sudip-Baldassia} (see~\cite{sudip-Nath} 
for a recent review on machine learning using quantum annealing). For a discussion on 
fault-tolerant quantum heuristics in the context of combinatorial optimizations, one may consult ref.~\cite{sudip-Sanders}, and ref.~\cite{sudip-Pelofske} for the indication of 
a major success using  parallel quantum annealing algorithms. For a discussion on computer aided  design of 
heterogeneous materials using quantum annealing algorithms, one may consult~\cite{sudip-Sahimi}. For recent 
extensive discussions on both theoretical and experimental developments, the readers are referred to the books~\cite{sudip-Tanaka,sudip-Dutta} and reviews~\cite{sudip-Albash,sudip-Hauke}.
\vspace{2mm}

\noindent {\em Acknowledgement}:  We are thankful to our colleagues Arunava Chakrabarti, Arnab Das and Purusattam Ray for their contributions at 
various stages of this development over last three decades, to Gabriel Aeppli, Amit Dutta, Uma Divakaran, Jun-ichi Inoue, Atanu Rajak, Thomas Rosenbaum, 
Diptiman Sen, Parongama Sen, Sei Suzuki, Ryo Tamura and Shu Tanaka for collaborations on this and closely related topics over the years, 
and to Elias Starchl for  his comments  on the manuscript  and help  with Fig.~\ref{entropy_plot}. BKC is grateful to the Indian Academy of Sciences for 
their support through a Senior Scientist Research Grant. SM thanks the SERB, DST (India) for partial financial support through the TARE scheme [file   no.:TAR/2021/000170] (2022).



\begin{thebibliography}{100} 
\bibitem{sudip-sk}
 D. Sherrington and S. Kirkpatrick, Phys. Rev. Lett. \textbf{35}, 1792 (1975).
 
\bibitem{sudip-SA}
 S. Kirkpatrick, C. D. Gelatt, and M. P. Vecchi, Optimization by Simulated Annealing,
Science, \textbf{220}, 671 (1983).
 
\bibitem{sudip-ray}
P. Ray, B. K. Chakrabarti  and A. Chakrabarti, Phys. Rev. B. \textbf{39}, 11828 (1989).

\bibitem{sudip-bikas}
S. Mukherjee, B. K. Chakrabarti, Eur. Phys. J. Special Topics \textbf{224}, 17 (2015).

\bibitem{sudip-Nishimori}
T. Kadowaki and H. Nishimori, Phys. Rev. E, \textbf{58}, 5355 (1998).

\bibitem{sudip-Farhi}
E. Farhi, J. Goldstone, S. Gutmann, J. Lapan, A. Ludgren and D. Preda, Science, \textbf{292}, 472 (2001).


\bibitem{sudip-Santoro}
G. E. Santoro, R. Marto{\v{n}}{\'a}k, E. Tosatti and R. Car,  Science, \textbf{295}, 2427 (2002).

\bibitem{sudip-Brooke}
J. Brooke, D. Bitko, T. F. Rosenbaum and G. Aeppli, Science \textbf{284}, 779 (1999).

\bibitem{sudip-das}
A. Das and B. K. Chakrabarti, Rev. Mod. Phys. \textbf{80}, 1061 (2008).

\bibitem{sudip-jsps}
S. Mukherjee and B. K. Chakrabarti, J. Phys. Soc. Jap., \textbf{88}, 061004 (2019).

\bibitem{sudip-Leschke}
 H. Leschke, C. Manai, R. Ruder and S. Warzel, Phys. Rev. Lett., \textbf{127}, 207204 (2021). 
 
\bibitem{sudip-Stachel}
E. Starchl and H. Ritsch, J. Phys. B: At. Mol. Opt. Phys. 55 (2022) 025501 (2022). 

\bibitem{sudip-Johnson}
M. W. Johnson, M. H. S. Amin, S. Gildert, T. Lanting, F. Hamze, N. Dickson, R. Harris, A. J. Berkley, J.  Johansson and P. Bunyk, Nature, 
\textbf{473}, 194 (2011).

\bibitem{sudip-Heim}
B. Heim, T. F. R{\o{}}nnow, S. V. Isakov and M. Troyer,  Science, \textbf{348}, 215 (2015). 

\bibitem{sudip-Baldassia}
C. Baldassia and R. Zecchinaa, Proc. Nat. Acad. Sc., \textbf{115}, 1457 (2018). 

\bibitem{sudip-Nath}
R. K. Nath, H. Thapliyal  T. S.  Humble, SN Comp. Sc., \textbf{2}, 365 (2021).

\bibitem{sudip-Sanders}
Y. R. Sanders, D.  W. Berry, P. C. S. Costa, L. W. Tessler, N. Wiebe,
C. Gidney, H. Neven, and R. Babbush, PRX Quantum \textbf{1}, 020312 (2020).


\bibitem{sudip-Pelofske}
E. Pelofske, G. Hahn,  H. N. Djidjev, Sci. Rep. \textbf{12}, 4499 (2022).


\bibitem{sudip-Sahimi}
M. Sahimi and P. Tahmasebi, Phys. Rep. \textbf{939}, 1 (2021). 

\bibitem{sudip-Dutta}
A. Dutta,  G. Aeppli, B. K. Chakrabarti,  U. Divakaran, T.
F. Rosenbaum and D. Sen, Quantum Phase Transitions in
Transverse Field Spin Models: From Statistical Physics to
Quantum Information,  Cambridge University Press,
Cambridge (2015).

\bibitem{sudip-Tanaka}
S.  Tanaka, R.  Tamura and  B. K Chakrabarti, Quantum spin glasses,
annealing and computation, Cambridge University Press, Cambridge (2017). 


\bibitem{sudip-Albash}
T. Albash and D. A. Lidar, Rev. Mod. Phys. \textbf{90}, 015002 (2018).

\bibitem{sudip-Hauke}
P. Hauke, H. G. Katzgraber, W. Lechner, H.
Nishimori and W. D. Oliver, Rep. Prog. Phys., \textbf{83}, 054401 (2020). 



 \end{thebibliography}
\end{document}